\documentclass[aps,pra,twocolumn,superscriptaddress]{revtex4-2}
\usepackage{amsmath,amssymb,graphicx,bbold}
\usepackage{graphicx}  % needed for figures
\usepackage{dcolumn}   % needed for some tables
\usepackage{bm}        % for math
\usepackage{color}

\usepackage{xcolor}
\usepackage{physics}
\usepackage{ulem}
\usepackage{float}
\usepackage{subfigure}

\begin{document}

\title{Spin to orbital angular momentum transfer in nonlinear wave mixing}

	\author{B. Pinheiro da Silva}
	\email{braianps@gmail.com}
\affiliation{Instituto de F\'{i}sica, Universidade Federal Fluminense, 24210-346 Niter\'{o}i, RJ, Brazil}

\author{W. T. Buono}
\affiliation{School of Physics, University of the Witwatersrand, Private Bag 3, Johannesburg 2050, South Africa}

\author{L. J. Pereira}
\affiliation{Instituto de F\'{i}sica, Universidade Federal Fluminense, 24210-346 Niter\'{o}i, RJ, Brazil}

\author{D. S. Tasca}
\affiliation{Instituto de F\'{i}sica, Universidade Federal Fluminense, 24210-346 Niter\'{o}i, RJ, Brazil}

\author{K. Dechoum}
\affiliation{Instituto de F\'{i}sica, Universidade Federal Fluminense, 24210-346 Niter\'{o}i, RJ, Brazil}

\author{A. Z. Khoury}
\email{azkhoury@id.uff.br}
\affiliation{Instituto de F\'{i}sica, Universidade Federal Fluminense, 24210-346 Niter\'{o}i, RJ, Brazil}
\date{\today}
\begin{abstract}

We demonstrate the spin to orbital angular momentum transfer in the nonlinear mixing of structured light beams. A vector vortex is coupled to a circularly polarized Gaussian beam in noncollinear second harmonic generation under type-II phase match. The second harmonic beam inherits the Hermite-Gaussian components of the vector vortex, however, the relative phase between them is determined by the polarization state of the Gaussian beam. This effect creates an interesting crosstalk between spin and orbital degrees of freedom, allowing the angular momentum transfer between them. Our experimental results match the theoretical predictions for the nonlinear optical response.
\end{abstract}
%\pacs{03.65.Vf, 03.67.Mn, 42.50.Dv}
%\vskip2pc 

\maketitle

\section{Introduction}

The interplay between spin and orbital angular momentum in nonlinear wave mixing has become an active research field. 
Many recent works have been devoted to the investigation of the crosstalk between different degrees of freedom in 
nonlinear processes. In our group we have investigated polarization controlled switching of orbital angular 
momentum (OAM) operations \cite{buono14,buono18}, radial-angular coupling in type-II second harmonic 
generation (SHG) \cite{buono17,buono20} and selection rules in optical parametric 
oscillation \cite{rafao,Rodrigues2018}. This subject has revealed a rich and fruitful research field with 
a much broader scope \cite{Zhi-Han2019,Fang2019,Zhi-Han2020,Zhi-Han2020-2}. The role played by different kinds 
of mode structures \cite{Sephton2019,Kumar2019,Pires2020,Pires2020-2,Rao2020} and controlled phase 
matching \cite{Chen2020} in parametric processes has been widely discussed. 
Beyond second harmonic generation, the nonlinear response to structured light fields has been also investigated 
in four-wave mixing \cite{Mallick2020,Benli2020,Sonja2021}, plasma \cite{Long2021}, surface science \cite{Dasgupta2019,Grigoriev2021} and magnetic structures \cite{Rouchon2021}. 
The generation of angular momentum supercontinuum in a ring array of coupled optical fibres 
has also been investigated \cite{Maitland2019}. The quantum optical description of the interplay between 
polarization and transverse mode structures in parametric down-conversion has been considered long ago \cite{Franca2001,Pardal2003}. More recently, this description was applied to parametric down-conversion 
of vector vortex beams both for spontaneous \cite{Khoury2020} and stimulated \cite{Walborn2020} processes. 
The quantum optical approach to the interaction between structured light and nonlinear has potential 
applications to novel quantum communication schemes \cite{Cai2018}. 

Much of these developments were made possible by the usage of 
new optical tools for shaping the phase and polarization distributions of a paraxial beam. Spatial light 
modulators (SLM) are nowadays a powerful tool for shaping the phase profile and control the diffraction 
properties of laser beams. They allow for easy and efficient generation of optical vortices, for example. 
Moreover, the development of specially fabricated plates, capable of shaping the polarization distribution 
of an optical beam, has made it possible to couple the spin and orbital angular momentum of a light beam. 
All these developments gave rise to the so called \textit{structured light}, a modern and important topic 
which has a number of applications in quantum information \cite{robert}, communication \cite{sid1, mehul2}, quantum cryptography \cite{mehul2, karimi}, optical tweezers \cite{miles11}, optical parametric oscillation \cite{rafao}, fiber 
optics \cite{sid2}, and many others \cite{dunlop2016,padgett17, shen19, forbes21}. 

Among these interesting structured light beams is the \textit{vector vortex} beam, a spin-orbit non-separable structure that 
is useful both in classical \cite{marcello, eutp} and quantum regimes. It can be used to study the crosstalk between 
spin and orbital angular momentum, as already was demonstrated in optical fiber systems \cite{sidgeo, sid3, sid4} and 
parametric down-conversion \cite{paulao}. In this work we demonstrate the spin-to-orbital angular momentum 
transfer in type-II second harmonic generation under noncollinear configuration. The concept of spin-orbit 
non-separable structures plays a central role as the spin-orbital angular momentum transfer is assisted by 
a vector vortex beam, which is nonlinearly mixed with a regular Gaussian beam prepared in an arbitrary polarization
state. 

\section{Spin-orbit coupling in nonlinear wave mixing}

This section describes the two-wave mixing process in noncollinear second harmonic generation under type-II phase match. Let us start by describing the incoming electric field of frequency $\omega$, which is compounded by two waves with different wave vectors $\textbf{k}_1$ and $\textbf{k}_2$. The corresponding electric field amplitudes are given by 
\begin{subequations}
\begin{eqnarray}
    \textbf{E}^{(\omega)}_{\textbf{k}_1} &=& [\mathcal{A}_{1x}(\textbf{r})\widehat{\textbf{x}} + \mathcal{A}_{1y}(\textbf{r})\widehat{\textbf{y}} ]e^{i\textbf{k}_1 \cdot \textbf{r}},
    \label{eq:input-1}\\
    \textbf{E}^{(\omega)}_{\textbf{k}_2} &=& (\alpha \widehat{\textbf{x}} + \beta \widehat{\textbf{y}}) \mathcal{A}_{2}(\textbf{r})e^{i\textbf{k}_2 \cdot \textbf{r}},
    \label{eq:input-2}
\end{eqnarray}
\end{subequations}
where $\mathcal{A}_{1x}(\textbf{r})$ and $\mathcal{A}_{1y}(\textbf{r})$ are the transverse structures carried by polarizations $x$ and $y\,$, respectively, of the beam $\textbf{k}_1$, and $\mathcal{A}_{2}(\textbf{r})$ is the structure of the beam $\textbf{k}_2$. The unit vectors $\widehat{\textbf{x}}$ and $\widehat{\textbf{y}}$ are, respectively, the horizontal and vertical polarization states, weighted by the complex numbers $\alpha$ and $\beta\,$, which obey the normalization relation $|\alpha|^2+|\beta|^2=1\,$.

\begin{figure}[h!]
	\includegraphics[width=1\columnwidth]{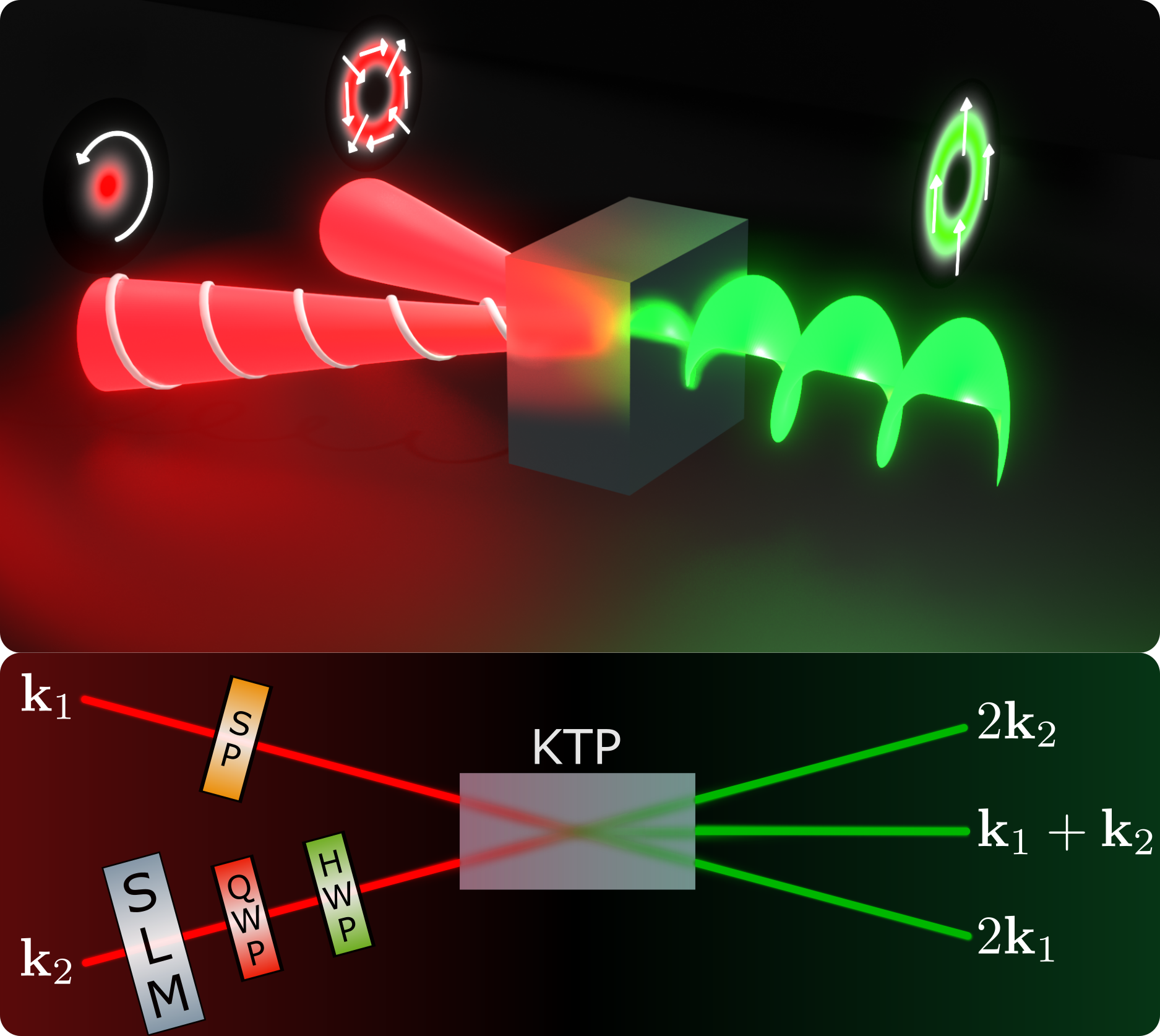}
	\caption{Experimental scheme for spin-orbit angular momentum transfer in second harmonic generation.}
	\label{fig:setup}
\end{figure}

Under noncollinear configuration, the output electric field of the second harmonic frequency $2\omega$ is formed by 
three contributions, each one associated with a different combination of the fundamental frequency components. These
contributions give rise to three different wave vectors at the second harmonic output: $2\textbf{k}_1$, $2\textbf{k}_2$, and $\textbf{k}_1 + \textbf{k}_2$ (shown in Fig. \ref{fig:setup}), which are associated with three simultaneous 
processes. Wave vectors $2\textbf{k}_1$ and $2\textbf{k}_2$ are associated with two independent frequency doubling 
processes of the incoming beams, while $\textbf{k}_1 + \textbf{k}_2$ corresponds to the nonlinear mixing of the 
input beams.
 \begin{figure*}
\centering
	\includegraphics[scale=0.25]{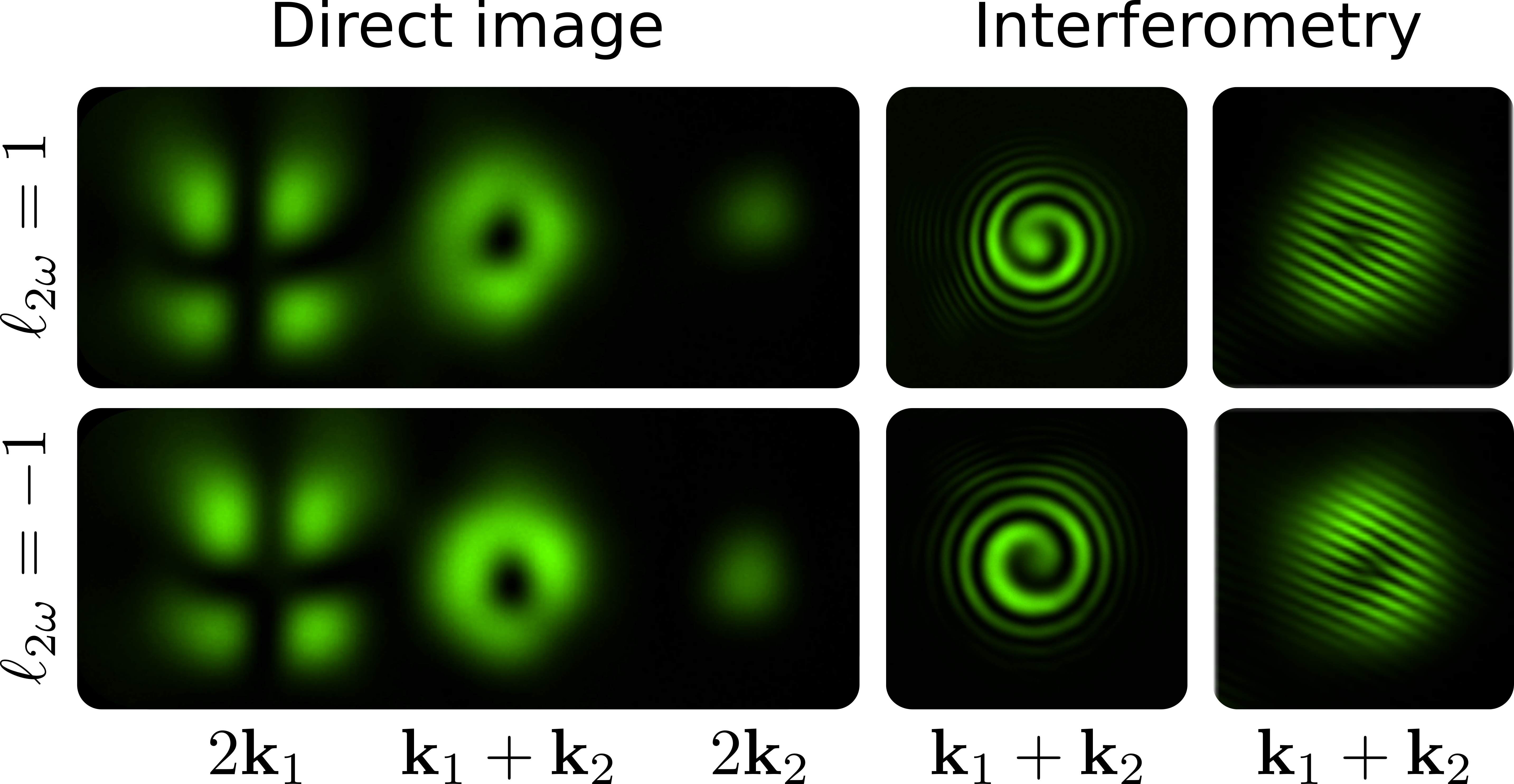}
	\caption{Images produced on the three outputs of the second harmonic field. Laguerre-Gaussian modes with topological charge $\ell=\pm 1$ are evident in the $\textbf{k}_1+\textbf{k}_2$ beam, as characterized by the spiral and forked intereference fringes shown on the right.}
	\label{fig:spirals}
\end{figure*}
In practice, we have three spatially resolved beams that exit the crystal along different directions, which facilitates the independent analysis of their transverse structures. Moreover, if the angle between $\textbf{k}_1$ and $\textbf{k}_2$ is small, we can neglect longitudinal components of the input electric fields along the different outgoing directions, as done in reference \cite{buono18}. The type-II phase match couples the horizontal and vertical polarization components of the fundamental field ($\omega$) to generate the second harmonic ($2\omega$) with vertical polarization. The resulting electric field amplitudes of the outgoing waves are 
\begin{subequations}
\begin{eqnarray}
\!\!\!\!\!\!\!\!\!\!\!\!\!\!\!\!\!\!\!\!\!\!\!\!
& &\textbf{E}^{(2\omega)}_{2\textbf{k}_1} = g_1 \mathcal{A}_{1x}(\textbf{r})\mathcal{A}_{1y}(\textbf{r}) e^{i2\textbf{k}_1\cdot \textbf{r}} \widehat{\textbf{y}}, \\
\!\!\!\!\!\!\!\!\!\!\!\!\!\!\!\!\!\!\!\!\!\!\!\!
& &\textbf{E}^{(2\omega)}_{2\textbf{k}_2}  = g_2\, \alpha\, \beta\, \mathcal{A}_{2}^2(\textbf{r}) e^{i2\textbf{k}_2\cdot \textbf{r}} \widehat{\textbf{y}}, \\
\!\!\!\!\!\!\!\!\!\!\!\!\!\!\!\!\!\!\!\!\!\!\!\!
& &\textbf{E}^{(2\omega)}_{\textbf{k}_1+\textbf{k}_2}  
        \!\!=\! g_{12}[\alpha \mathcal{A}_{1y}(\textbf{r}) + \beta \mathcal{A}_{1x}(\textbf{r})] \mathcal{A}_{2}(\textbf{r}) e^{i(\textbf{k}_1 + \textbf{k}_2)\cdot \textbf{r}} \widehat{\textbf{y}},
\end{eqnarray}
\end{subequations}
where $g_1,\,g_2$ and $g_{12}$ are the coupling coefficients which are proportional to the nonlinear susceptibility of the medium. Here it is important to note that the output wave along $\textbf{k}_1+\textbf{k}_2$ carries two contributions. 
The first one comes from the coupling between the $y$ polarization of the input wave $\textbf{k}_1$ with the $x$ 
polarization of the wave $\textbf{k}_2\,.$ This term is proportional to the product 
$\alpha \mathcal{A}_{1y} \mathcal{A}_{2}\,$.  
The second contribution comes from the coupling between the $x$ polarization of the input wave $\textbf{k}_1$ with the $y$ 
polarization of the wave $\textbf{k}_2$ and is proportional to the product 
$\beta \mathcal{A}_{1x} \mathcal{A}_{2}\,$. 
Therefore, the resulting transverse structure at the output wave is a superposition composed by the transverse 
modes of input wave $\textbf{k}_1\,$, weighted by the polarization coefficients of the input $\textbf{k}_2\,$. 
Provided the transverse structure $\mathcal{A}_{2}$ is simply a Gaussian mode, it will only produce a rescaling 
of the output waist. As we will see, this spin-orbit crosstalk allows the transfer of the spin angular momentum of the input 
wave $\textbf{k}_2$ to the orbital angular momentum of the second harmonic output $\textbf{k}_1 + \textbf{k}_2\,$.

\section{Spin-to-orbital angular momentum transfer}

The spin-orbit angular momentum transfer becomes evident when we write the input and output modes in terms 
of Laguerre-Gaussian (LG) and Hermite-Gaussian (HG) functions. 
The first one is the solution of the paraxial wave equation in cylindrical coordinates. The LG function reads 
\begin{align}
&\textrm{LG}_{\ell,p}(\tilde{r},\theta) = \mathcal{R}_{\ell,p}(\tilde{r})\,
\tilde{r}^{\vert\ell\vert} \, e^{i\ell\theta}\,,
%\label{eq:lg}
\nonumber
\\
&\mathcal{R}_{\ell,p}(\tilde{r}) = \frac{\mathcal{N}_{\ell p}}{w}\,
\textrm{L}_p^{\vert\ell\vert}\left(\tilde{r}^2\right) \,e^{-\frac{\tilde{r}^2}{2}}\,e^{-i\Phi_N}\,,\label{eq:lg}
\\
&\Phi_N = \frac{k\,\tilde{r}^2}{2R} + (N+1) \arctan\left( z/z_0\right)\,,\quad \tilde{r} = \sqrt{2}\, r/w\,,
\nonumber
\end{align}
where $N=2p+\vert\ell\vert$ is the mode order, $\ell$ is the topological charge, $p$ the radial order, $\textrm{L}_p^{\vert{\ell}\vert}$ are generalized Laguerre polynomials and $\mathcal{N}_{\ell p}$ is a normalization constant. The beam parameters are the wave-front radius $R$, the width $w$ and the Rayleigh length $z_0\,$. These modes can carry orbital angular momentum of $\ell\hbar$ per photon.

The solutions in Cartesian coordinates are the Hermite-Gaussian modes, which are given by
\begin{align}
&\textrm{HG}_{m,n}(\tilde{x},\tilde{y}) = \frac{\mathcal{N}_{mn}}{w}\,
\textrm{H}_m\left(\tilde{x}\right) \,\textrm{H}_n\left(\tilde{y}\right) \,e^{-\frac{\tilde{x}^2 + \tilde{y}^2}{2}} 
\,e^{-i\Phi_N}\,,
\nonumber\\
&\tilde{x} = \sqrt{2}\, x/w\,,\quad \tilde{y} = \sqrt{2}\, y/w\,,
\end{align}
where $\mathcal{N}_{mn}$ is the proper normalization constant, $\textrm{H}_n$ are the Hermite polynomials with index $n$, and the HG mode order is $N=m+n\,$. These modes do not have orbital angular momentum. Both the  LG  and  HG  modes constitute orthonormal and complete bases of the transverse mode vector space.  In this sense, it is possible to decompose any LG mode in terms of HG modes of the same order, and the opposite is also true \cite{holandeses}.

Our reasoning about the spin-orbit angular momentum transfer becomes straightforward when we highlight some 
multiplicative properties of the HG modes as follows 
\begin{subequations}
\begin{eqnarray}
    \textrm{HG}_{0,0}(\tilde{\textbf{r}})\,\textrm{HG}_{m,n}(\tilde{\textbf{r}}) &\propto&  \textrm{HG}_{m,n}(\sqrt{2}\tilde{\textbf{r}})\,, \label{eq:hgmult-a}\\
    \textrm{HG}_{m,0}(\tilde{\textbf{r}})\,\textrm{HG}_{0,n}(\tilde{\textbf{r}}) &\propto&  \textrm{HG}_{m,n}(\sqrt{2}\tilde{\textbf{r}})\,.\label{eq:hgmult-b}
\end{eqnarray}
\end{subequations}
As we can see, apart from a rescaling of the transverse coordinates, these HG products result in new HG 
modes that combine the properties of the factor modes.
The spin to orbital angular momentum transfer is achieved when specific structures are prepared in the incoming fields with the fundamental frequency $\omega\,$. 
Let us assume that the input beam with wave vector $\textbf{k}_1$ is prepared in a vector vortex structure corresponding to $\mathcal{A}_{1x}(\tilde{\textbf{r}}) = \textrm{HG}_{0,1}(\tilde{\textbf{r}})$ and $\mathcal{A}_{1y}(\tilde{\textbf{r}})= \textrm{HG}_{1,0}(\tilde{\textbf{r}})$. The input beam  with wave vector $\textbf{k}_2$ is assumed to be in a Gaussian mode, 
that is $\mathcal{A}_2(\tilde{\textbf{r}}) = \textrm{HG}_{0,0}(\tilde{\textbf{r}})\,$, with an arbitrary polarization 
$\alpha\widehat{\textbf{x}} + \beta \widehat{\textbf{y}}\,$. 
Using the HG product properties given by Eqs. \eqref{eq:hgmult-a}-\eqref{eq:hgmult-b} for this configuration, the transverse structure of each output beam in the second harmonic frequency will be 
\begin{subequations}
\begin{align}
    \mathcal{B}_{1}(\tilde{\textbf{r}}) &= \mathcal{A}_{1x}(\textbf{r})\mathcal{A}_{1y}(\textbf{r})
    =\textrm{HG}_{1,1}(\sqrt{2}\tilde{\textbf{r}})\,,\\
    \mathcal{B}_{2}(\tilde{\textbf{r}}) &= \alpha\beta \mathcal{A}_{2}^{2}(\textbf{r})
    =\alpha \beta\, \textrm{HG}_{0,0}(\sqrt{2}\tilde{\textbf{r}})\,, \\
    \mathcal{B}_{12}(\tilde{\textbf{r}}) &= 
    \left[\alpha \mathcal{A}_{1y}(\textbf{r}) + \beta \mathcal{A}_{1x}(\textbf{r})\right]\mathcal{A}_{2}(\textbf{r})
    \nonumber\\
    &= \alpha \, \textrm{HG}_{1,0}(\sqrt{2}\tilde{\textbf{r}}) + \beta \, \textrm{HG}_{0,1}(\sqrt{2}\tilde{\textbf{r}})\,.\label{eq:tout12}
\end{align}
\end{subequations}
By inspecting these equations, we conclude the following: 
\\
\begin{enumerate}
    \item The transverse structure $\mathcal{B}_{1}(\tilde{\textbf{r}})$ on output $2\textbf{k}_1$ carries a 
    fixed $\textrm{HG}_{1,1}$ mode.
    \item The transverse structure $\mathcal{B}_{2}(\tilde{\textbf{r}})$ on output $2\textbf{k}_2$ carries a 
    Gaussian mode $\textrm{HG}_{0,0}$ with its amplitude affected by the product $\alpha\beta$ between its polarization coefficients.
    \item The transverse structure $\mathcal{B}_{12}(\tilde{\textbf{r}})$ on output $\textbf{k}_1 + \textbf{k}_2$ 
    inherits the HG components of the vector vortex beam incoming on 
    $\textbf{k}_1\,$, and the complex weights $\alpha$ and $\beta$ from the polarization of the input beam $\textbf{k}_2\,$.
\end{enumerate}

Therefore, by changing these complex weights, it is possible to create any first order transverse mode in the second harmonic. 
For example, when the $\textbf{k}_2$ input beam is circularly polarized ($\beta= \pm i\alpha$), it carries a spin angular momentum (SAM) of $S\hbar$ per photon, where $S=\pm 1$. Using these weights in the Eq. (\ref{eq:tout12}), the transverse structure $\mathcal{B}_{12}(\tilde{\textbf{r}})$ becomes a $\textrm{LG}_{\pm 1,0}(\tilde{\textbf{r}})$, which carries OAM of $\pm \hbar$ per photon. It means that the spin of the input beam $\textbf{k}_2$ was transferred to the orbital angular momentum of the output beam $\textbf{k}_1 +\textbf{k}_2$ due the nonlinear wave mixing process.

\section{Experimental results}

The angular momentum transfer from the spin of the input beam $\textbf{k}_2$ to the orbital angular momentum of 
the second harmonic can be shown experimentally with the scheme illustrated in Fig. \ref{fig:setup}. The source of the input beams is an infrared Nd:Yag laser ($\lambda=1064nm$). On $\textbf{k}_1$, we produce a vector vortex beam using an s-plate (SP). The transverse structure and the polarization of $\textbf{k}_2$ are prepared with a spatial light modulator (SLM) followed by a sequence of a quarter- (QWP) and a half-wave (HWP) plate. The two input beams pass through a potassium 
titanyl phosphate (KTP) crystal cut for type II phase match, where the nonlinear wave mixing process occurs. The results of the second harmonic generation ($\lambda=532nm$) are the three output beams $2\textbf{k}_1$, $2\textbf{k}_2$, and $\textbf{k}_1 + \textbf{k}_2\,$.

First, we set the polarization of the input beam $\textbf{k}_2$ to right circular ($\beta=i\alpha$). In this case the output beam $\textbf{k}_1 + \textbf{k}_2$ acquires a topological charge $\ell_{2\omega}=1$ ($\textrm{LG}_{1,0}$) and carries orbital angular momentum inherited from the spin ($S=1$) of the input beam $\textbf{k}_2$, as illustrated at the top of the Fig. \ref{fig:setup}. Then, we switch to left circular polarization ($\beta=-i\alpha$, $S=-1$), in which case the orbital angular momentum of the $\textbf{k}_1 + \textbf{k}_2$ output is also switched to $\ell_{2\omega}=-1$. The corresponding experimental results are shown in Fig. \ref{fig:spirals}. The first column is the direct image of the three outputs beams: $2\textbf{k}_1$, $\textbf{k}_1 + \textbf{k}_2$, and $2\textbf{k}_2$. The orbital angular momentum of the Laguerre modes is evinced by two different types of interferometry. The first one is the spherical wave interference between the Gaussian mode on $2\textbf{k}_2$ and the Laguerre-Gaussian mode on $\textbf{k}_1 + \textbf{k}_2\,$. In this case, the OAM is determined by the number and the sense of the spiral fringes. The second measurement employs plane wave interference, in which the orbital angular momentum is given by the fork fringes. The results of Fig. \ref{fig:spirals} confirm the theoretical predictions. The imperfections in the intensity distribution of the modes are mainly due to the walk-off effect present in the wave plates and the nonlinear crystal. Other sources of imperfections are related to limited precision in mode and polarization preparation, and due to small overlaps between the three output beams.

\begin{figure}[h!]
\centering
	\includegraphics[scale=0.4]{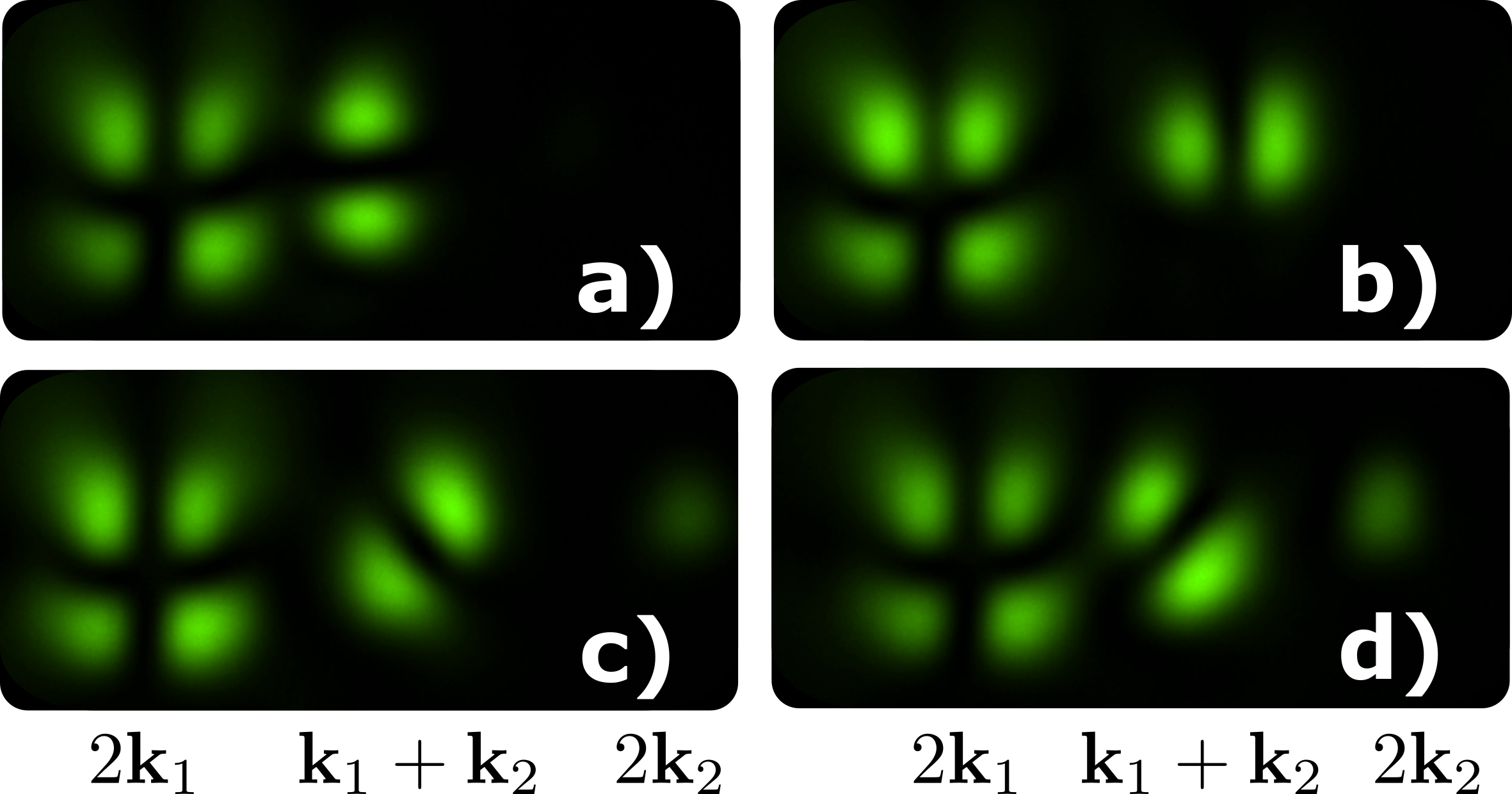}
	\caption{Hermite Gaussian modes produced on $\textbf{k}_1 + \textbf{k}_2$ by different orientations of the linear polarization prepared on $\textbf{k}_2\,$. }
	\label{fig:hg}
\end{figure}

\begin{figure*}[ht!]
\centering
	\includegraphics[width=1.8\columnwidth]{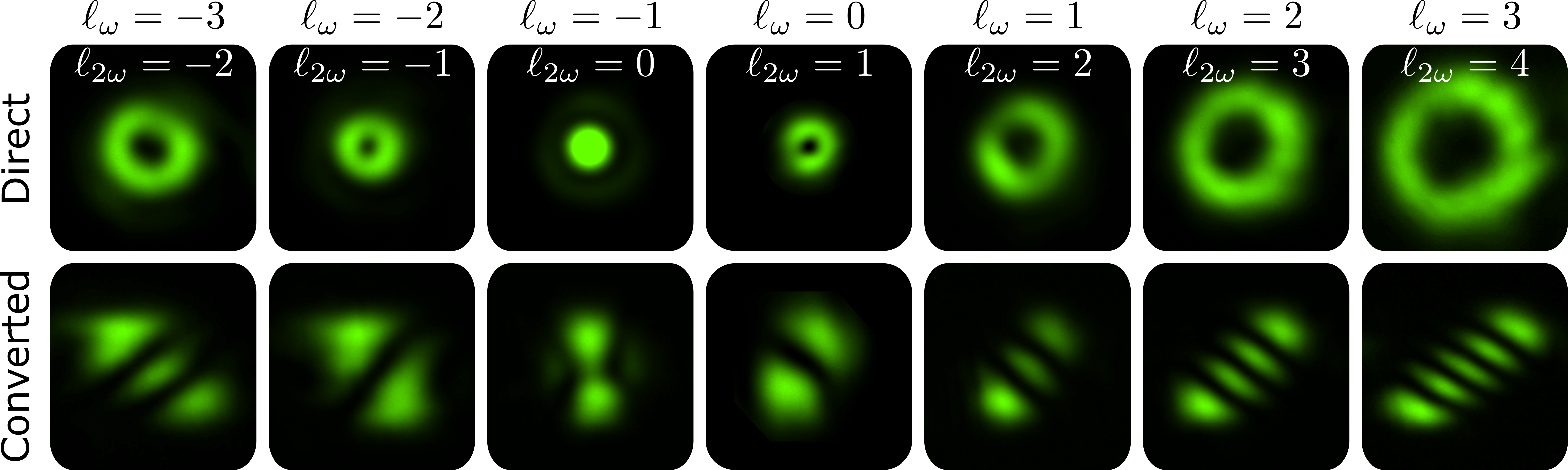}
	\caption{Spin and orbital angular momentum transfer in second harmonic generation with a right circularly polarized 
	$\textbf{k}_2$ beam ($S=+1$), carrying topological charges $-3\leq\ell_{\omega}\leq +3$ and $p=0\,$. 
	Top row displays the direct image acquisition. Bottom row displays the tilted lens conversion for 
	easy identification of the second harmonic topological charge.}
	\label{fig:p0}
\end{figure*}

\begin{figure}[H]
\centering
	\includegraphics[scale=0.4]{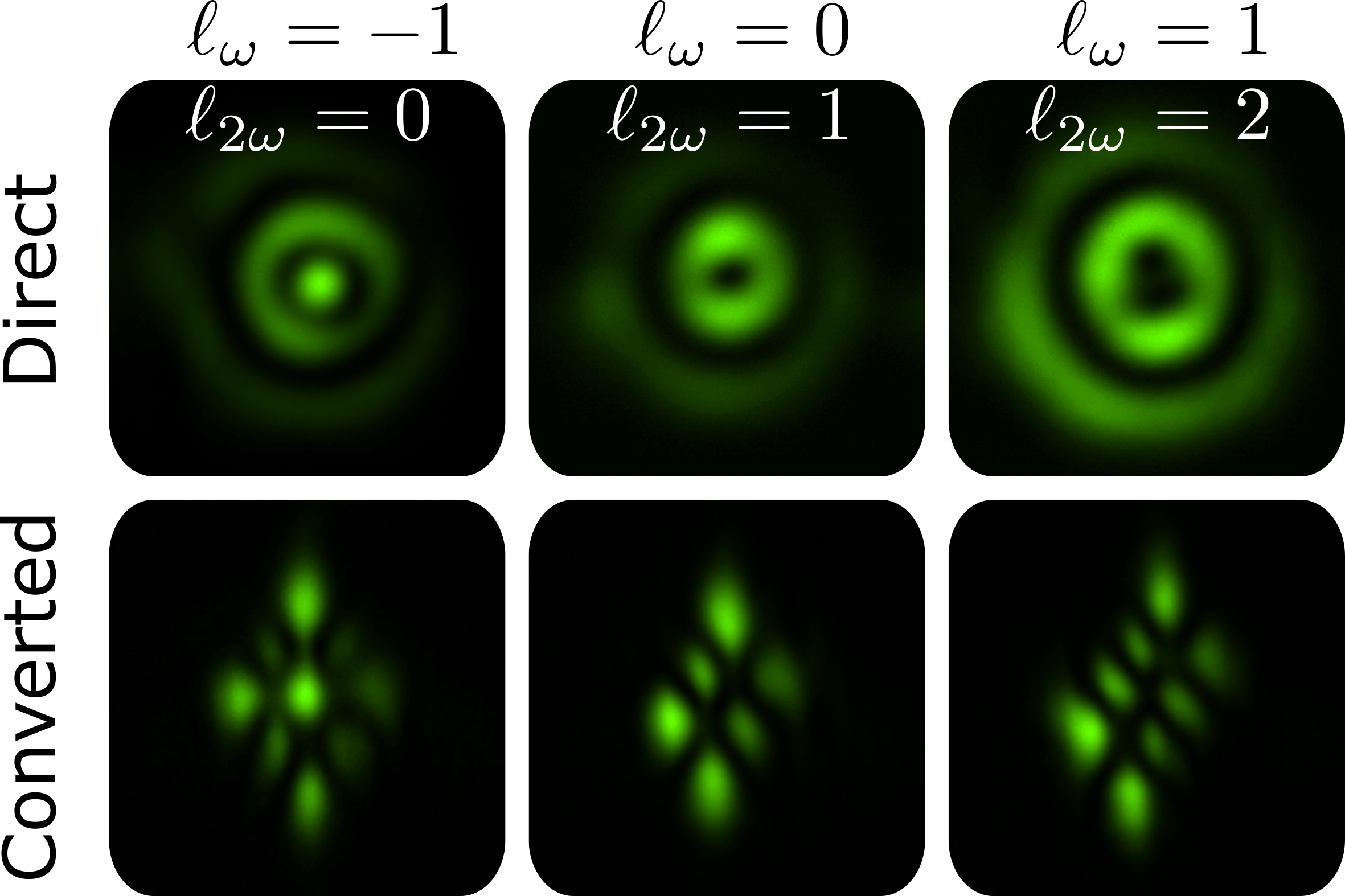}
	\caption{Same as Fig. \ref{fig:p0} for input beam $\textbf{k}_2$ prepared in Laguerre-Gaussian modes with $-1 \leq \ell_\omega \leq 1$ and $p=1\,$.}
	\label{fig:p1}
\end{figure}

We have also tested the transfer of four different linear polarization states: vertical ($\alpha=0$, $\beta=1$), horizontal ($\alpha=1$, $\beta=0$), diagonal ($\beta=\alpha$), and anti-diagonal ($\beta=-\alpha$). The experimental results are shown in Figs. \ref{fig:hg}(a-d), respectively. The orientations of the Hermite-Gaussian modes produced on $\textbf{k}_1 + \textbf{k}_2$ are: a) $88^\circ \pm 4 ^\circ$, b) $1^\circ  \pm 4 ^\circ$, c) $51^\circ  \pm 4 ^\circ$, and d) $-49^\circ  \pm 4 ^\circ$. They are in good agreement with the theoretical predictions. As expected, when the polarization of the input beam $\textbf{k}_2$ is either horizontal or vertical, the intensity of the beam $2\textbf{k}_2$ vanishes because $\vert \alpha\beta\vert^2=0\,$. The small difference between the theoretical and experimental angles, as well as the mode imperfections, have the same origins as before. 

If the input beam $\textbf{k}_2$ carries orbital and spin angular momentum, the output beam 
$\textbf{k}_1 + \textbf{k}_2$ will inherit both contributions purely as OAM. 
In order to demonstrate this, we prepared the input beam $\textbf{k}_2$ with right circular polarization ($S = +1$) and transverse mode $\textrm{LG}_{\ell_\omega,p_\omega}$ with topological charge $\ell_\omega\,$. The input beam $\textbf{k}_1$ was kept in the same vector vortex structure as before. For this configuration the transverse structure of the second harmonic output $\textbf{k}_1 + \textbf{k}_2$ becomes
\begin{eqnarray}\label{eq:lgs}
    \mathcal{B}_{12}(\tilde{\textbf{r}}) &=& \textrm{LG}_{1,0}(\tilde{\textbf{r}}) \, \textrm{LG}_{\ell_\omega,p_\omega}(\tilde{\textbf{r}}) \\ \nonumber
    &=&\mathcal{R}_{1,0}(\tilde{r})\,\mathcal{R}_{\ell_\omega,p_\omega}(\tilde{r}) \, \tilde{r}^{\vert \ell_\omega \vert +1} \, e^{i(\ell_\omega+1)\theta},
\end{eqnarray}
where $\mathcal{R}_{\ell,p}(\tilde{r})$ was defined in Eq. \eqref{eq:lg}. The resulting topological charge  
is $\ell_{2\omega}=\ell_\omega+1\,$, which corresponds to the added spin and orbital angular momentum of the input 
beam $\textbf{k}_1\,$. Note that if $\vert \ell_\omega \vert +1 \neq \vert \ell_\omega+1\vert$, then radial rings 
appear second harmonic as a consequence of the radial-angular mismatch discussed in Ref. \cite{buono20}.

%%%%%%%% era aqui

We obtained two sets of experimental results that confirm the predictions of Eq. \eqref{eq:lgs}. In the first one, 
the input beam $\textbf{k}_2$ is prepared in seven different Laguerre-Gaussian modes with $-3 \leq \ell_\omega \leq 3$ and $p=0\,$. The resulting images are shown in Fig. \ref{fig:p0}. The top images were acquired directly and the 
bottom ones were registered after a tilted lens mode converter that allows for easy identification of the 
resulting orbital angular momentum and radial order \cite{tilted, eu19, eu21}. The tilted lens analysis is 
explained in Appendix A. For the cases where the input spin and orbital angular momentum have opposite 
signs ($\ell_\omega < 0$), a radial-angular mismatched beam is formed in the second harmonic and the 
consequent radial orders are manifested in the weak outer rings of the far-field images.
In the second set of experimental results, we prepared three different Laguerre-Gaussian modes with 
$-1 \leq \ell_\omega \leq 1$ and radial order $p=1\,$. The corresponding results are shown in Fig. \ref{fig:p1}. 
All topological charges measured in both sets of experimental results confirm the theoretical prediction of 
Eq. \eqref{eq:lgs}.

\section{Conclusion}

We demonstrated an interesting effect of spin-to-orbital angular momentum transfer in nonlinear mixing of structured light beams. First, we show that an arbitrary polarization state of an input Gaussian beam is transferred to the second 
harmonic frequency when it is mixed with a vector vortex input. The second harmonic beam inherits the Hermite-Gaussian 
components of the vector vortex, weighted by the polarization coefficients of the Gaussian input. This spin-orbit 
crosstalk is caused by the type-II phase match condition that couples orthogonal electric field components of 
the interacting beams. When the Gaussian input beam is replaced by a Laguerre-Gaussian mode carrying a topological 
charge, both spin and orbital angular momentum are transferred to the second harmonic output, verifying the general 
relation $\ell_{2\omega} = S + \ell_{\omega}\,$. Our experimental results are in good agreement with
the theoretical calculation of the second harmonic field amplitudes. This interesting effect can be useful for 
classical and quantum information transfer between different wavelengths, mediated by spin-orbit coupling. 
The potential developments in the quantum domain are under investigation.

\section*{Acknowledgments}
The authors acknowledge financial support from the Brazilian Agencies, Conselho Nacional de Desenvolvimento 
Tecnol\'ogico (CNPq), Funda\c c\~{a}o Carlos Chagas Filho de Amparo \`{a} Pesquisa do Estado do Rio de Janeiro 
(FAPERJ), Coordena\c c\~{a}o de Aperfei\c coamento de Pessoal de N\'ivel Superior (CAPES - Finance Code 001) 
and the Brazilian National Institute of Science and Technology of Quantum Information (INCT-IQ 465469/2014-0).

\bibliographystyle{unsrt}
\bibliography{bibfile}
\clearpage
\appendix
\section{Mode analysis with a tilted lens}

\begin{figure}[h!]
\centering
	\includegraphics[width=1\columnwidth]{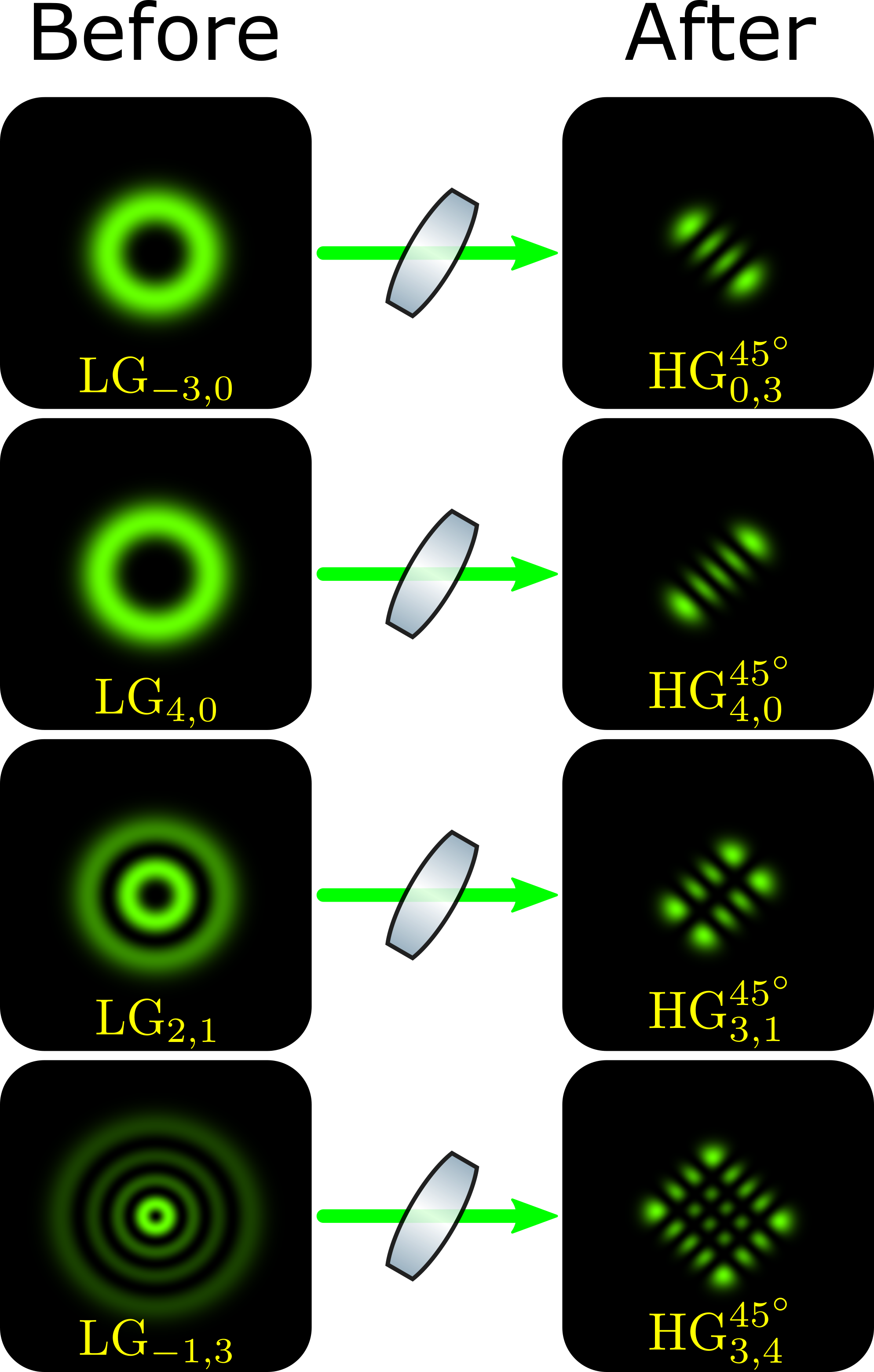}
	\caption{Transverse structure before and after the tilted lens.}
	\label{fig:tl}
\end{figure}

The astigmatic transformation implemented by a simple spherical lens with a small tilt angle allows for easy 
identification of the OAM ($\ell$) and the radial order ($p$) of pure Laguerre-Gaussian modes \cite{tilted}. 
Moreover, it can be used to recognize OAM superpositions \cite{eu19, eu21}. 
This method is trustworthy and easy to apply to an efficient analysis of orbital angular momentum. 
An incident Laguerre-Gaussian mode $\textrm{LG}_{\ell,p}$ undergoes an 
astigmatic transformation through the tilted lens and near the focal plane it is converted to 
a rotated Hermite-Gaussian mode $\textrm{HG}_{m,n}^{45^\circ}$, where
\begin{equation}
\textrm{HG}_{m,n}^{45^\circ} \equiv 
\textrm{HG}_{m,n} \left(\frac{x+y}{\sqrt{2}},\frac{x-y}{\sqrt{2}}\right)\,.
\end{equation}
This transformation is illustrated in Fig. \ref{fig:tl}.
The orbital angular momentum and radial index of the incident mode are readily determined from
\begin{enumerate}
    \item $\ell = m-n\,$.

    \item $p=\mathrm{min}(m,n)\,$.
\end{enumerate}

\clearpage

\end{document}